\documentclass[12pt,preprint]{aastex}
\usepackage{epsfig,wrapfig,graphicx}

\def\kms{km~s$^{-1}$}
\def\ms{m~s$^{-1}$}
\def\mjup{M$_{\rm Jup}$}

\def\msun{M$_{\odot}$}
\def\mpl{0.82}
\def\kamp{22}
\def\star{GJ\,849}
\def\peryear{5.16}
\def\perday{1890}
\def\ecc{0.06}
\def\fe{[Fe/H]}
\def\arel{2.35}
\def\msini{$M_P\sin i~$}
\def\farcs{\mbox{\ensuremath{.\!\!^{\prime\prime}}}}%  %
\begin{document}
\title{A Long--period Jupiter--mass Planet Orbiting the Nearby M~Dwarf \star}

\author{R. Paul Butler\altaffilmark{2}, 
John Asher Johnson\altaffilmark{3},
Geoffrey W. Marcy\altaffilmark{3},
Jason T. Wright\altaffilmark{3},
Steven S. Vogt\altaffilmark{5},
Debra A. Fischer\altaffilmark{4}}

\email{paul@dtm.ciw.edu}

\altaffiltext{1}{ Based on observations obtained at the
W.M. Keck Observatory, which is operated jointly by the
University of California and the California Institute of
Technology.  Keck time has been granted by both NASA and
the University of California.}

\altaffiltext{2}{Department of Terrestrial Magnetism, Carnegie Institution
of Washington, 5241 Broad Branch Road NW, Washington D.C. USA 20015-1305}

\altaffiltext{3}{Department of Astronomy, University of California,
Berkeley, CA USA  94720}

\altaffiltext{4}{Department of Physics and Astronomy,
San Francisco State University, San Francisco, CA, USA 94132}

\altaffiltext{5}{UCO/Lick Observatory, University of California
at Santa Cruz, Santa Cruz CA USA 95064}

\begin{abstract}
We report precise Doppler measurements of \star\ (M3.5V) that reveal
the presence of a planet with a minimum mass of \mpl~\mjup\ in a
\peryear~year orbit. At $a = \arel$~AU, \star\,b is the 
first Doppler--detected planet discovered around an M~dwarf orbiting
beyond 0.21~AU, and is only the second Jupiter--mass planet discovered
around a star less massive than 0.5~\msun. 
This detection brings to 4 the number of M~stars known to
harbor planets. Based on the results of our survey of 1300 FGKM
main--sequence stars we
find that giant planets within 2.5~AU are $\sim$3 times more
common around GK stars than around M stars. Due to the \star's
proximity of 8.8~pc, the planet's 
angular separation is 0\farcs27, making this system a prime
target for high--resolution imaging using adaptive optics and future
space--borne missions such as the \emph{Space Interferometry
  Mission}. We also find evidence of a linear trend in the velocity time
series, which may be indicative of an additional planetary
companion. 
\end{abstract}


\section{Introduction}

Of the 152 stars within 200~pc of the Sun known to harbor
planets\footnote{For the updated catalog of extrasolar
  planet, their parameters, and the properties of the host stars see
  http://exoplanets.org.}, the majority are Sun--like, with
masses between 0.7 and 1.3~\msun\ \citep{butler06}. Main--sequence
stars with masses greater 
than 1.3~\msun\ (spectral types earlier than F8V) are typically
unsuitable for precision radial velocity (RV) monitoring because their
spectra lack narrow absorption lines \citep{galland05, johnson06b} and
excessive astmospheric ``jitter'' \citep{wright05}. However, at the
lower end of the mass spectrum, M--type dwarfs are much more 
amenable to precision Doppler measurements, with the primary
observational limitation being their relative faintness. Over 200
M~dwarfs ($M_* < 0.6$~\msun) have been monitored by various
Doppler surveys using large telescopes \citep[e.g.][]{wright04b, kurster03,
  endl03}. These radial velocity planet searches have so far
discovered five planets orbiting 
only three host stars: the triple system around GJ\,876
\citep{marcy98,delfosse98,marcy01,rivera05}, and the Neptune--mass 
planetary companions to GJ\,436 \citep{butler04} and GJ\,581
\citep{bonfils05b}. Only one of these three systems, GJ\,876, 
contains Jupiter--mass planets, and despite the $\gtrsim 2$~yr
duration of these surveys, none has revealed a planet beyond
0.21~AU. \footnote{Three additional low--mass planet host stars have been
discovered by gravitational lensing surveys \citep{bond04, gould06,
  beaulieu06}. However, due to the faintness of these candidates, the
stellar mass estimates of all but one have large uncertainties. Using
Hubble ACS imaging, \citet{bennett06} determined that
OGLE--2003--BLG--235 is likely a late K-type dwarf, with a mass of
0.6~\msun.}

The RV precision attainable from the spectra
of middle--age ($> 2$~Gyr) M~dwarfs is similar to that of G-- and
K--type stars, and the stars themselves typically exhibit low levels
of photospheric 
jitter \citep{wright05}. Additionally, the Doppler reflex amplitude
scales as $K \propto a^{-1/2} M_P M_*^{-1/2}$, which makes 
planets of a given mass easier to detect around low--mass stars. The
detectability of  
planets orbiting M dwarfs is therefore comparable to that of FGK
stars, allowing a comparative understanding of the planet formation
process in different stellar mass regimes.
Based on the lack of planet detections in their 
survey of 90 M~dwarfs, \citet{endl06} estimate that fewer than 1.27\%
of stars with $M_* < 0.6$~\msun\ harbor Jovian mass planets with $a <
1$~AU, which stands in stark contrast to the 5\% occurrence rate of
gas giants around solar--type stars \citep{marcy05a}.
This finding seems to indicate that protoplanetary disks around
low--mass stars produce Jovian planets at a decreased rate compared to
the disks of Sun--like stars. 

The formation of planets around low--mass stars has been studied in the
context of the core accretion planet 
formation model. In this formation scenario, rocky cores are built up
through collisions in the protoplanetary disks around young stars
\citep{wetherill89, kokubo01}. Once a critical core mass is reached,
gas accumulates onto the core through a run--away accretion process,
resulting in a gas giant by the time the supply of disk gas is exhausted
\citep[e.g.][]{pollack96}. \citet{laughlin04} showed that the lower disk
masses, decreased surface density of solids and longer orbital
time scales of M~dwarf protoplanetary disks inhibit the growth of  
planetesimals enough such that the disk gas dissipates before the 
critical core mass is reached. The resulting prediction is that there
should be a relative abundance of
Neptune--mass ``ice giants'' around M~dwarfs, but a far smaller
number of gas giants. This prediction agrees well with the findings of
\citet{ida05b}, who studied the frequency of planets for stellar
masses ranging from 0.4 to 1.5~\msun. Based on their Monte Carlo
simulations, they find the number of giant planets drops
significantly with decreasing mass for $M_* < 1$~\msun. These
theoretical results are in accordance with the available observational data.

However, due to the relatively shorter time baselines of most M~dwarf
Doppler surveys, the current observational data only provide
information about planets orbiting within $\sim 2$~AU of their
host stars. As the
durations of the M~dwarf planet surveys increase, it will become
evident whether the current observed paucity of gas giants holds for
larger orbital separations, or if there exists a separate, larger
population of Jupiter--mass planets residing in long--period
orbits. Additionally, the prediction of inhibited core growth around
lower mass stars is made under the assumption that disk mass scales
proportionally with stellar mass \citep{laughlin04}. If this
assumption is relaxed, \citet{kornet06} find that low--mass stars
actually form giant planets at an \emph{increased} rate compared to
Solar--mass stars. Searching for planets at larger orbital separations
will provide an important test of these theories.

We report the detection of a Jupiter--mass planet in a \peryear~yr
orbit around the M3.5 dwarf, \star. We present the stellar
characteristics of the host star in \S~\ref{stellar}. In
\S~\ref{orbit} we discuss our observations and orbit solution.
We conclude in \S~\ref{discussion} with a discussion of the latest
M~dwarf planetary system and the occurrence of planets around M~dwarfs.

\section{Stellar Properties}
\label{stellar}

\begin{figure}[h!]
\epsscale{1}
\plotone{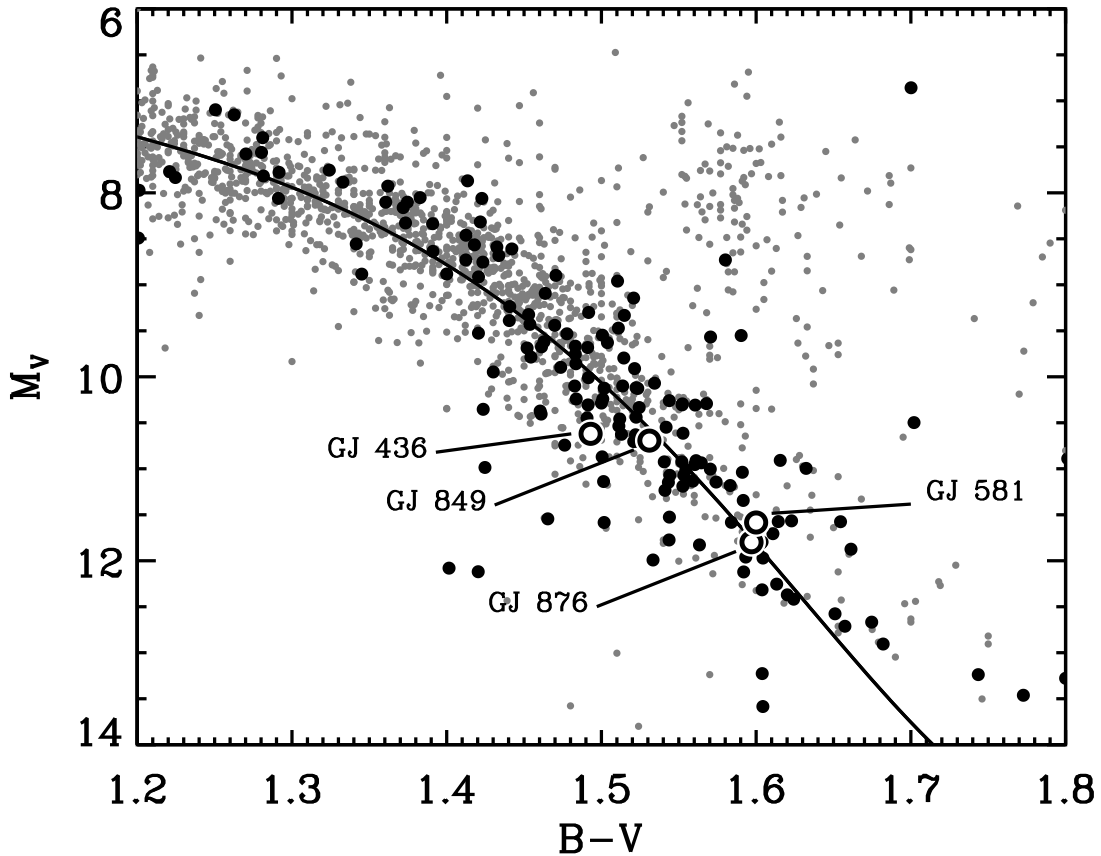}
\figcaption{\footnotesize H-R diagram illustrating the properties of the four
  M~dwarfs known to harbor planets (large open circles) compared to:
  the sample of stars in the  California \& California Planet Search
  (black filled circles); the stars in the Hipparcos catalog with
  distances $<  50$~pc (gray filled circles); and the mean Hipparcos
  main--sequence described by \citet{wright05} (solid line). \label{mstar_hr}} 
\end{figure}

We have been monitoring a sample of 147 low--mass, late--K through M dwarfs
as part of the California and Carnegie Planet Search
\citep[CCPS;][]{butler06, rauscher06}. One of them, \star\ (HIP\,109388,
LHS\,517), is an M3.5V star with $V=10.42$ and $B-V = 1.52$ (ESA
1997). Its Hipparcos--based parallax ($\pi = 114$~mas) implies a
distance of 8.8~pc and an absolute visual magnitude $M_V =
10.69$. Figure \ref{mstar_hr} shows the position of \star\ with
respect to the other M--type stars with known planetary companions. Also
shown is the CCPS sample, the stars  listed in the Hipparcos catalog
within 50~pc of the Sun, and the mean Hipparcos main--sequence as
defined by \citet{wright05}.  

Despite the star's location slightly below the mean Hipparcos main--sequence,
the K--band photometric metallicity--luminosity calibration of
\citet{bonfils05a} and IR magnitudes of \citet{leggett92} suggest
that \star\ has a metallicity consistent with solar: \fe$ = +0.16 \pm
0.2$. The K--band 
mass--luminosity calibration of \citet{delfosse00} yields a 
stellar mass $M_* = 0.47 \pm 0.04$~\msun. This mass
estimate agrees well with the $0.51 \pm 0.05$~\msun\ mass
predicted by the K--band mass--luminosity relationship of
\citet{henry93}. We adopt the mean of these two estimates, 
$M_* = 0.49 \pm 0.05$~\msun, as the mass of \star. 

Examination of our high--resolution spectra reveals no Balmer line
emission. \citet{delfosse98} report a projected equatorial
rotational velocity $V_{rot}\sin{i} = 2.4$~\kms, and \citet{marcy92} measure
$V_{rot}\sin{i} = 1.0 \pm 0.6$~\kms. The low
chromospheric activity and slow rotation of \star\ are consistent with a
middle--age dwarf older than 3~Gyr (Andrew West, private communication).

\section{Observations and Orbital Solution}
\label{orbit}

\begin{wrapfigure}{r}[0.2cm]{9cm}
\includegraphics[width=8cm, height=10cm]{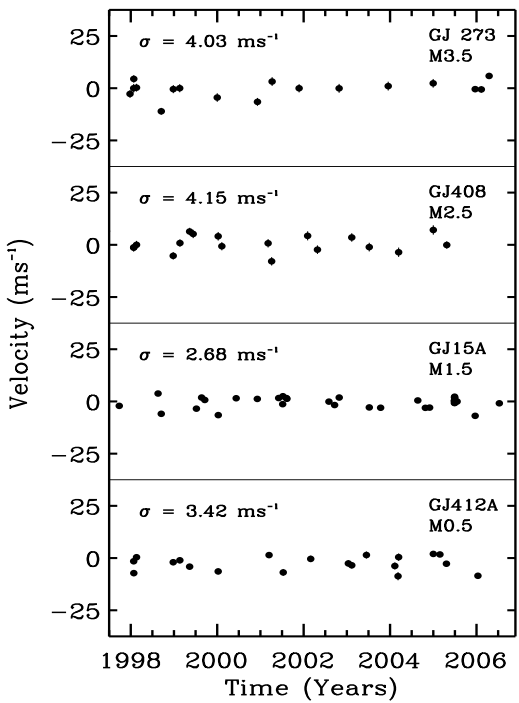}
\caption{\footnotesize Radial velocity time series for four stable M~dwarfs in
  our Keck Doppler survey. \label{standards}} 
\end{wrapfigure}

We have been monitoring GJ 849 with the Keck I 10\,m telescope
for 6.9 years as part of the NASA Keck M~Dwarf Survey and the
California and Carnegie Planet Search (CCPS).  We obtained
high--resolution spectra using the HIRES echelle spectrometer
\citep{vogt94} with an Iodine cell mounted directly
in front of the entrance slit \citep{valenti94}. The Doppler shift
is measured from each star--plus--iodine observation using the
modeling procedure described by \citet{butler96}. Figure
\ref{standards} shows our velocity measurements for four stable 
M~Dwarfs, demonstrating our long--term Doppler precision of 3-4~\ms.

\begin{figure}[bh!]
\epsscale{1.0}
\plotone{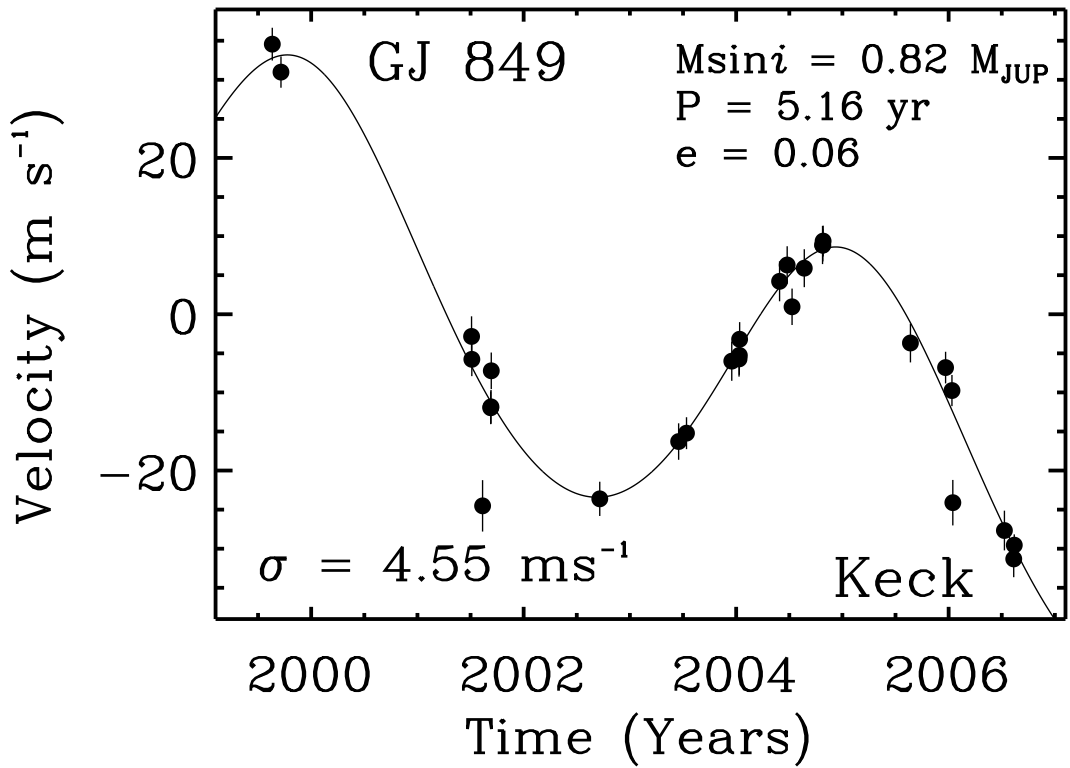}
\caption{\footnotesize Doppler velocities for GJ 849.  These data span 6.9~years.
The best--fit Keplerian, including a linear -4.6~\ms~year$^{-1}$
trend, is shown as a solid line. The orbital solution has a period,
$P = \peryear$~yr, semiamplitude, $K = 22$~\ms, eccentricity $e =
0.06$, yielding \msini$ = 0.82$~\mjup\ and semimajor axis of 2.35~AU.
The RMS to the is 4.55~\ms.  } 
\label{vel_plot}
\end{figure}

A total of 29 precision Doppler measurements of \star\ spanning 6.9
years are listed in Table~\ref{vel_table} and shown in
Figure~\ref{vel_plot}. The measurement uncertainties listed in
Table~\ref{vel_table} represent the weighted standard deviation of the
velocities measured from the 700 2~\AA wide spectral ``chunks'' 
used in our Doppler analysis \citep{butler96}. The solid line shows
the best--fit Keplerian  
plus linear trend, which has a slope of -4.6~\ms\,yr$^{-1}$.
The Keplerian parameters are listed in Table~\ref{kep_pars}, along
with their estimated uncertainties, which were derived using a Monte
Carlo method \citep[e.g.][]{marcy05b}. Our best fit orbital solution
yields a \peryear\ year period, 
velocity semi--amplitude $K = \kamp$~\ms, and eccentricity 
$\ecc \pm 0.09$---consistent with circular. Using our adopted stellar
mass $M_* = 0.49$~\msun, we calculate a minimum planet mass $M_P\sin{i} =
\mpl$~\mjup\ and semimajor axis $a =2.35$~AU. The RMS of the fit
residuals is 4.55~\ms, resulting in a reduced $\chi_{\nu}^2 = 1.6$.

Figure~\ref{vel_plot} shows that much of the RMS scatter is dominated
by two observations that sit more
than 2$\sigma$ below the best--fit Keplerian. These two observations
have the largest measurement uncertainties in our data set, and if
they are excluded the RMS of the best--fit Keplerian plus linear trend
improves to  2.1~\ms\ and $\chi_{\nu}^2 = 0.91$. Exclusion of
these outliers does not change the derived orbital parameters beyond
uncertainties listed in Table~\ref{kep_pars}. We see no correlations
or additional periodicities in the residuals.

\section{Discussion}
\label{discussion}

We present here the detection of a Jupiter--mass planetary companion
to the M3.5 dwarf \star. This detection brings to 4 the number of
M~dwarfs harboring Doppler--detected planets, together with GJ\,876
\citep{marcy01, rivera05}, GJ\,436 \citep{butler04}, and GJ\,581
\citep{bonfils05b}. The \star\ planetary system is remarkable in
two respects: the orbital separation ($a = \arel$~AU) of the 
planet is more than an order of magnitude greater than any other
Doppler--detected M~dwarf planet, and the system is only the second
known to include a Jupiter--mass planet.

We have been monitoring 147 late K and M~dwarfs ($0.2 \leq M_* <
0.6$~\msun) at Keck Observatory for nearly 7 years, with a typical
Doppler precision of 3~\ms\ \citep{wright05, rauscher06}. Of these
stars, 114 have 8 or more observations spanning a minimum time
baseline of 5.5 years. Within this subset of M~dwarfs
our survey is sensitive to planets that induce $K > 12$~\ms\
(4-$\sigma$ level) for periods P$<$ 3.5 yr. These limits on $P$ and
$K$  correspond to minimum planet masses of \msini $>$ 0.4~\mjup\ and
orbital separations of $a < $ 1.8 AU, assuming a nominal stellar mass
of 0.5 \msun.  Only one star in this sample, GJ\,876, 
harbors planets that meet these criteria. Thus the occurrence of
planets having a  minimum mass over 0.4 \mjup\ within 1.8 AU around M
dwarfs is $\sim$0.9\%, albeit with large fractional uncertainty 
($\sigma \approx 1$~\%). Note that GJ 849b is not 
included in this domain of $a$ and \msini\ as it has $a > $ 1.8 AU. If
we extend the maximum orbital separation from 1.8 to 2.5 AU, then GJ
849b is included.  This relaxed threshold corresponds to $K > 10$~\ms\ 
implying only a 3-$\sigma$ detection threshold.   Thus for $a <$ 2.5
AU, the occurrence rate of giant mass planets is 2/114 =
1.8$\pm$1.2\%, but remains  uncertain due to small--number
statistics. 

The planet occurrence rate for M~dwarfs ($M_* < 0.5$~\msun) can be
compared to the corresponding rate for higher--mass G-- and K--type
stars observed 
at Keck as part of the CCPS. In this sample, there are 232 GK
stars with $0.6 \leq M_* \leq 1.1$~\msun\ and 8 or more
observations spanning more than 4 years. Of these stars,
13 have jupiter mass planets within 1.8~AU, yielding an occurrence rate of
$5.6 \pm 1.6$\%. Thus, giant mass planets 
are almost 6 times more likely to be detected orbiting within 1.8~AU
of GK stars than around M~dwarfs. The fraction of planets orbiting GK stars
within 2.5~AU is also $5.6 \pm 1.6$\%, resulting in a factor of 3 
higher likelihood of finding planets orbiting solar--mass stars
compared to M~dwarfs. 

This simple analysis does not account for the sparse nominal sampling rate
(8 observations spanning 4-6 years), which may miss signals with
amplitudes near $K = 12$~\ms, especially those in highly eccentric
orbits. Indeed, several of 
our M~dwarfs show RV variations consistent with Jupiter mass companions, but
lack sampling sufficient enough to determine a unique orbital
solution. Thus, additional monitoring is necessary before firm
conclusions can be drawn between the fraction of M~dwarf planet hosts
compared to higher--mass stars. 

Also not addressed by our analysis is the effect of metallicity, which
has already been established as a strong tracer of planet occurrence
\citep{fischer05b}. If there exists any correlation between mass
and metallicity within our overall stellar sample, then the apparent
relationship between stellar mass and planet occurrence would be
difficult to separate from the effects of metallicity. Such a
correlation between mass and metallicity could arise as a result of
systematic errors in LTE abundance determinations or selection biases
in our stellar sample. While no such selection bias affects our sample
of M dwarfs---which is complete for distances less than 20~pc and
apparent magnitudes brighter than 11.5---it is possible that a bias
exists at the high--mass end  of our sample. Accounting for such
effects is beyond the scope of this paper and will be addressed in a
future  publication (Johnson et al. 2007, in preparation).

At 8.8~pc, the orbital separation of \star\,b corresponds
to a projected separation of 0\farcs27. Thus, the proximity of \star\
provides a unique opportunity for high--resolution imaging using
adaptive optics and future space--borne astrometric missions such as
the \emph{Space Interferometry Mission}. 

\acknowledgements 

We acknowledge support by NSF grant AST-9988087, NASA grant
NAG5-12182, and travel support from the Carnegie Institution
of Washington (to RPB), NASA grant NAG5-8299 and NSF grant
AST95-20443 (to GWM), and NSF for its grants AST-0307493 (to SSV). We
thank NASA and the University of California 
for their allocations of Keck telescope time.  This research has made use of
the Simbad database operated at CDS, Strasbourg France, and the
NASA ADS database.  Finally the authors wish to extend thanks
to those of Hawaiian ancestry on whose sacred mountain of
Mauna Kea we are privileged to guests.  Without their generous
hospitality, the Keck observations presented herein would not
have been possible.

%\bibliographystyle{apj}
%\bibliography{apj-jour,myrefs}

\begin{deluxetable}{rrr}
\tablecaption{Velocities for GJ 849}
\label{vel_table}
\tablewidth{0pt}
\tablehead{
\colhead{JD}           &    \colhead{RV}         & \colhead{error} \\
\colhead{($-$2450000)}   &  \colhead{(m s$^{-1}$)} & \colhead{(m s$^{-1}$)} 
}
\startdata
  1410.0215  &    38.9  &  2.0 \\
  1439.8654  &    35.3  &  2.0 \\
  2095.0814  &     1.5  &  2.6 \\
  2096.0458  &    -1.4  &  2.2 \\
  2133.0128  &   -20.2  &  3.3 \\
  2160.9092  &    -7.6  &  2.1 \\
  2161.8459  &    -7.5  &  2.1 \\
  2162.8870  &    -2.9  &  2.3 \\
  2535.8516  &   -19.3  &  2.2 \\
  2807.0106  &   -12.0  &  2.4 \\
  2834.0130  &   -10.9  &  2.1 \\
  2989.7201  &    -1.7  &  2.6 \\
  3014.7104  &    -1.4  &  2.2 \\
  3015.7110  &    -0.9  &  2.4 \\
  3016.7060  &     1.1  &  2.2 \\
  3154.0798  &     8.5  &  2.6 \\
  3180.1084  &    10.6  &  2.4 \\
  3196.9314  &     5.3  &  2.3 \\
  3238.9290  &    10.7  &  2.4 \\
  3301.8384  &    13.6  &  2.4 \\
  3302.7425  &    13.3  &  2.1 \\
  3303.7984  &    14.0  &  2.0 \\
  3603.9387  &     1.5  &  2.4 \\
  3724.7115  &    -1.8  &  2.1 \\
  3746.7182  &    -6.0  &  2.0 \\
  3749.6979  &   -19.9  &  2.9 \\
  3927.0148  &   -22.9  &  2.5 \\
  3959.0867  &   -24.5  &  2.3 \\
  3960.9584  &   -22.7  &  1.4 \\
\enddata
\end{deluxetable}

\begin{deluxetable}{lcc}
\tablecaption{Orbital Solution and Stellar Properties for \star
\label{kep_pars}}
\tablewidth{0pt}
\tablehead{
\colhead{Parameter} & \colhead{Value} & \colhead{Uncertainty}
}
\startdata
Orbital period $P$ (days) &  \perday & 130 \\
Orbital period $P$ (years) &  \peryear & 0.35 \\
Velocity semiamplitude $K$ (m\,s$^{-1}$) & \kamp  & 2 \\
Eccentricity $e$ & \ecc & 0.09\\
Periastron date (Julian Date) & 2451462  & 540 \\
Linear Velocity Trend (\ms\,year$^{-1}$) & -4.6 & 0.8 \\
$\omega$ (degrees) & 351 & 60 \\
M$\sin i$ (\mjup) & \mpl \\
semimajor axis (AU) & \arel \\ 
N$_{\rm obs}$ & 29 \\
RMS (m\,s$^{-1}$) & 4.55 \\
\enddata
\end{deluxetable}

\end{document}